\documentclass[11pt]{iopart}
\usepackage{bm}
\usepackage{hyperref}
\usepackage{amssymb}
\usepackage{amsopn}

\newcommand{\be}{\begin{equation}}
\newcommand{\ee}{\end{equation}}
\newcommand{\bea}{\begin{eqnarray}}
\newcommand{\eea}{\end{eqnarray}}

\def\be{\begin{equation}}
\def\ee{\end{equation}}
\def\bea{\begin{eqnarray}}
\def\eea{\end{eqnarray}}
\def\a{\alpha}
\def\b{\beta}

\def\a{\alpha}

\begin{document}
\hfill DAMTP-2007-12

\title{Non-linear vector perturbations in a contracting universe}
\date{\today}
\author{Filipe C. Mena$^\flat$, David J. Mulryne$^\sharp$ and Reza Tavakol$^\star$}
\address{$^\flat$ Departamento de Matem\'{a}tica, Universidade do Minho, 4710 Braga, Portugal}
\address{$^\sharp$ DAMTP, Centre for Mathematical Sciences, University of
  Cambridge, Wilberforce Road, Cambridge, CB3 0WA,
  United Kingdom}
\address{$^\star$ Astronomy Unit, School of Mathematical
Sciences,
Queen Mary, University of London, Mile End Road, London, E1 4NS,
United Kingdom}
\eads{\mailto{fmena@math.uminho.pt}, \mailto{d.mulryne@damtp.cam.ac.uk}, \mailto{r.tavakol@qmul.ac.uk}}

\begin{abstract}
A number of scalar field models proposed as alternatives to the
standard inflationary scenario involve contracting phases which
precede the universe's present phase of expansion. An important
question concerning such models is whether there are effects which
could potentially distinguish them from purely expanding
cosmologies. Vector perturbations have recently been considered in
this context. At first order such perturbations are not supported by
a scalar field. In this paper, therefore, we consider second order
vector perturbations. We show that such perturbations are generated
by first order scalar mode-mode couplings, and give an explicit
expression for them. We compare the magnitude of vector
perturbations produced in collapsing models with the corresponding
amplitudes produced during inflation, using a number of suitable
power--law solutions to model the inflationary and collapsing
scenarios. We conclude that the ratios of the magnitudes of these
perturbations depend on the details of the collapsing scenario as
well as on how the hot big bang is recovered, but for certain cases
could be large, growing with the duration of the collapse.
\end{abstract}

\pacs{04.25.Nx, 98.80.Hw}

%\submitto{JCAP}

\maketitle

%%%%%%%%%%%%%%%%%%%%%%%%%%%%%%%%%%%%%%%%%%%%%%%%%%%%%%%%%%%%%%%%%%%%%
\section{Introduction}
%%%%%%%%%%%%%%%%%%%%%%%%%%%%%%%%%%%%%%%%%%%%%%%%%%%%%%%%%%%%%%%%%%%
Recent years have witnessed tremendous advances in observational
cosmology. High--precision data from observations of the Cosmic
Microwave Background (CMB) and high redshift surveys have provided
strong evidence for a nearly spatially--flat universe with a
primordial spectrum of adiabatic, Gaussian and nearly
scale--invariant density perturbations, in excellent agreement with
the predictions of the simplest inflationary models
\cite{Peiris:2003ff,Spergel:2006hy}. Despite these important
successes, however, crucial questions remain. Paramount among these
is whether the inflationary scenario can be embedded within a
 fundamental theory. While attempts in this direction continue,
 a number of alternative scenarios have been put forward
motivated by string/M-theory. Among these are the pre-big bang
\cite{Gasperini:2002bn,Lidsey:1999mc} and the ekpyrotic/cyclic
scenarios
\cite{Khoury:2001wf,Steinhardt:2001st,Steinhardt:2002ih,Khoury:2003rt,Steinhardt:2004gk}.
Despite their differences, an important ingredient shared by these
models is the existence of a contracting epoch preceding a poorly
understood bounce into the present expansionary phase of the
universe. Interestingly, there are now many proposals for realising
bouncing cosmologies within the context of string/M-theory
\cite{Biswas:2005qr, Aref'eva:2007uk}, braneworld models \cite{Shtanov:2002mb,
Burgess:2003tz, Kanti:2003bx, Hovdebo:2003ug, Foffa:2003gt} and
recently Loop Quantum Cosmology \cite{Singh:2003au, Lidsey:2004ef,
Ashtekar:2006rx, Singh:2006im}, which add weight to the feasibility
of scenarios of this type. Thus, given the successes of inflationary
cosmology, on the one hand, and the importance of considering
alternative scenarios, on the other,
%embeddable within
%candidate theories of quantum gravity
it is important to seek observational signatures that could
distinguish between these alternative possibilities.

An important tool in this connection is provided by cosmological
perturbation theory. There are in general three types of
perturbations, namely scalar, vector and tensor, which decouple at
the linear level. A great deal of work has been done on studying the
evolution of such perturbations in an expanding universe, both at
the linear (for a review see \cite{Mukhanov:1990me}) and nonlinear
levels (see e.g. \cite{NH, Nakamura,Bartolo}). A number of studies
have also been made of the evolution of perturbations in collapsing
models. In order to succeed as an alternative to inflation, any
model of the early universe must provide an explanation of the
primordial scalar perturbations, which are the seeds of structure in
the universe. The amplitude and spectrum of these scalar
perturbations are highly constrained by CMB observations, and
although problems remain, both the pre-big bang and the
cyclic/ekpyrotic scenarios have the potential to explain their
origin. In this sense scalar perturbations may not be enough to
distinguish between these scenarios.

The next step is to consider other perturbation modes. Tensor
perturbations are important as they can in principle be observed as
a primordial gravitational wave background and can thus provide a
way of distinguishing between different scenarios. Indeed, it has
been shown that the cyclic/ekpyrotic scenario has very different
predictions for tensor perturbations when compared to single field
inflationary models, with the cyclic scenario producing a negligible
level of gravitational waves with a blue tilted spectrum
\cite{Boyle:2003km}.

Until recently, however, little consideration was given to the
evolution of vector perturbations. This is because these
perturbations are generally assumed not to be important in the
inflationary scenario for two reasons. First, in an expanding FLRW
universe first order metric vector perturbations decay and hence
rapidly become insignificant. Secondly, once inflation has begun all
matter except for the scalar field driving inflation is rapidly red
shifted, with the result that the universe is effectively sourced
solely by scalar field matter. This implies that vector perturbations
play a minimal role since, as we shall see explicitly in the next section,
a scalar field does not support vector perturbations at first order.

In a collapsing scenario, however, the situation is potentially very
different, as has been discussed recently in an interesting study
\cite{brand1}. Indeed, the first objection is no longer valid, since
during a collapsing phase first order vector perturbations grow and
in principal this growth could have important observational
consequences for the collapsing scenarios \cite{brand1}. The second
objection is, however, still valid for the collapsing scenarios
mentioned above, as these are also sourced solely by scalar
field matter. Noting this, the authors of \cite{brand1}
proceeded by considering vector perturbations in presence of
pressureless dust. However, the more natural setting to consider
perturbations in this context remains that of scalar
fields, where first order vector perturbations are absent. In view
of this, it is important to ask whether nonlinear vector
perturbations can be supported in such regimes, and how they evolve
during collapsing phases.  This is the aim of the present paper.

We shall study the evolution of second order vector perturbations in
a collapsing universe sourced by a scalar field. An important
generic feature of nonlinear perturbations is that vector, tensor
and scalar modes couple. Consequently, we expect second order vector
perturbations to be produced, for example, by first order scalar
mode-mode couplings. The analogous production of second order tensor
perturbations has recently been studied \cite{ACW,OPDUC}. It is
possible, therefore, that vector perturbations could provide a
signature of a collapsing phase which would be absent or highly
suppressed in an expanding universe. Indeed, the possibility that
second order vector perturbations could act as seeds for large scale
cosmic magnetic fields has previously been considered
\cite{Matarrese-etal04,Maartens}, as has their contribution to the
polarisation of the CMB \cite{Mollerach-etal03}.

The structure of the paper is as follows. In Section \ref{back} we
introduce the flat FLRW model in the presence of a scalar field, and
consider perturbations about this background; subsection
\ref{firstord} reviews first order scalar and vector perturbations,
and an expression for the second order vector perturbations is
derived in  subsection \ref{secondord}. In Section \ref{analytic} we
consider a scalar field self--interacting through an exponential
potential, which allows a power--law solution to the cosmological
evolution. By considering different power--law regimes we are able
to study concrete examples of interest analytically, and calculate
and compare the amplitude of vector perturbations in each case in
Section \ref{Bcalc}.
Finally, Section \ref{conclusion} contains our discussions and
conclusions.

Throughout lower case Latin indices take values $0,1,2,3$ and Greek
indices $1,2,3$.
%%%%%%%%%%%%%%%%%%%%%%%%%%%%%%%%%%%%%%%%%%%%%%%%%%%%%%%%%%%%%%%%%%%%%%%%%%%%%%%
\section{Background model and Perturbations}
\label{back}
%%%%%%%%%%%%%%%%%%%%%%%%%%%%%%%%%%%%%%%%%%%%%%%%%%%%%%%%%%%%%%%%%%%%%%%%%%%%%%%%
We consider as our background model a flat
Friedmann-Lemaitre-Robertson-Walker (FLRW) metric in the form \be
\label{flat-RW} ds^2=g^{(0)}_{ab} dx^a dx^b =
a(\tau)^2(-d\tau^2+\delta_{\alpha\beta}dx^\alpha dx^\beta)\,, \ee where
$\tau$ is the conformal time which is related to the coordinate time
$t$ through $dt=ad\tau$. We take the universe to be sourced by a
scalar field $\phi$ with a stress energy tensor given by \be
\label{Tab-scalar}
T_{ab}=\phi_{,a}\phi_{,b}-\frac{1}{2}g_{ab}\phi^{,c}\phi_{,c}-g_{ab}V(\phi)\,,
\ee where $V( \phi )$ is the associated scalar field potential. The
background evolution equations are given by the Friedmann equation
\be \label{Fried} H^2 = \frac{8\pi G}{3}\left
(\frac{\dot{\phi}^2}{2} + V(\phi) \right ) \,, \ee where $H=\dot
a/a$ is the Hubble parameter and a dot denotes differentiation with
respect to the coordinate time $t$,
%; the $\dot{H}$ equation \be
%\label{Hdot} \dot{H}=-{4\pi G \dot{\phi}^2} \,, \ee
and the
Klein-Gordon equation for the scalar field \be \label{scalar}
\ddot{\phi}+3H\dot{\phi}+\frac{\partial V(\phi)}{\partial \phi}=0\,.
\ee Throughout we shall use the formalism developed in \cite{NH} in
order to give the vector perturbations up to second order in a flat
FLRW universe sourced by a scalar field.

In order to derive the perturbation equations we recall that
the perturbed FLRW metric can, up to second order,
be written in the usual form \cite{NH}
\bea
g_{00}&=&-a^2\left(1+2(A^{(1)}+A^{(2)})\right)\,, \\
g_{0 \alpha} &=& -a^2 (B^{(1)}_{\alpha}+B^{(2)}_{\alpha})\,,\\
g_{\a\b}&=& a^2 \left ( g^{(0)}_{\alpha \beta}+2(C^{(1)}_{\alpha
\beta}+C^{(2)}_{\alpha \beta}) \right )\,, \eea where $g^{(0)}_{\alpha
\beta}$ is the background 3-metric and the superscripts $(1)$ and
$(2)$ denote first and second order quantities, respectively. The
perturbation variables can be decomposed, at each order $i$, as
\bea
B^{(i)}_\alpha &\equiv& \beta^{(i)}_{,\alpha} + B^{(vi)}_{\alpha}\,, \\
C^{(i)}_{\alpha \beta} &\equiv& \varphi^{(i)} g_{\alpha \beta} +
\gamma^{(i)}_{,\alpha|\beta}+C^{(vi)}_{(\alpha|\beta)}+C^{(ti)}_{\alpha \beta}
\eea
with $\partial^\alpha B_\alpha^{(vi)}=\partial^\alpha C_\alpha^{(vi)}=
\partial^\alpha C^{(ti)}_{\alpha\beta}=C^{\alpha(ti)}_{~\alpha}=0$. In this splitting, the variables $\varphi^{(i)}$ and $\gamma^{(i)}$
represent scalar perturbations, while vector and tensor
perturbations are denoted by the superscripts $(v)$ and $(t)$,
respectively.

To proceed we require a gauge, which we choose to be the Poisson
gauge, defined by $\beta^{(i)}=\gamma^{(i)} =C^{(vi)}_{\alpha} =0$.
This gauge is a generalisation of the longitudinal gauge to include
vector and tensor modes.

We also note that we do not use the energy frame here. Instead, we
follow \cite{NH} and use the normal frame, in which {the scalar}
field stress-energy tensor can be identified with the fluid
stress-energy tensor in such a way that the velocity perturbations
are related to the scalar field perturbations by \cite{NH}
\begin{eqnarray}
\label{vector-identif}
v^{(1)}_\alpha & = & -\frac{1}{a\dot\phi}\phi^{(1)}_{,\alpha}, \\
\label{vector-identif2}
v^{(2)}_\alpha & = &
-\frac{1}{a\dot\phi}\left ( \phi^{(2)}_{,\alpha}+\phi^{(1)}_{,\alpha}(\dot\phi^{(1)}+\dot\phi\varphi^{(1)})
\right)\,.
\end{eqnarray}
We shall decompose the velocity perturbations as
\[
v^{(i)}_\alpha=\partial_\alpha v^{(i)}+v_\alpha^{(vi)},
\]
with $\partial^\alpha v^{(i)}_\alpha=0$. We shall assume
$C_{\alpha\beta}^{(t1)}=0$ throughout and, in the next section, we
shall show that in our scalar field background we must also have
$v^{(v1)}_{\alpha}=B_{\alpha}^{(v1)}=0$.
%%%%%%%%%%%%%%%%%%%%%%%%%%%%%%%%%%%%%%%%%%%%%%%%%%%%%%%%%%%%%%%%%%%%%%%%%%
\subsection{First order perturbations}
\label{firstord}
%%%%%%%%%%%%%%%%%%%%%%%%%%%%%%%%%%%%%%%%%%%%%%%%%%%%%%%%%%%%%%%%%%%%%%%%%%
The first order perturbation equations in the present model have
been considered extensively by many authors. An important feature of
such perturbations is that the evolution equations for
scalar, vector and tensor perturbations decouple and hence can be
studied separately.

Confining ourselves to the Poisson gauge, the metric perturbations
$A^{(1)}$ and scalar perturbations $\varphi^{(1)}$ satisfy the
relation $A^{(1)}=-\varphi^{(1)}$ and their evolution equation becomes
\be \label{FOS1} \ddot{\varphi}^{(1)}+  \left (H
-2\frac{\ddot{\phi}}{\dot{\phi}}\right ) \dot{\varphi}^{(1)} -
\frac{\nabla^2\varphi^{(1)}}{a^2} + 2\left
(\dot{H}-H\frac{\ddot{\phi}}{\phi}\right)\varphi^{(1)} = 0\,. \ee
Employing the field equations together with the above
fluid--scalar-field identification, to first order, the scalar field
perturbation can also be given in terms of $\varphi^{(1)}$ as \be
\label{FOS2} 4\pi G
\dot{\phi}\phi^{(1)}=-(H\varphi^{(1)}+\dot\varphi^{(1)})\,.
\ee
While
considering first order scalar perturbations, we take the
opportunity to define the curvature perturbations on comoving
hypersurfaces \cite{Bardeen}
\be
\label{Curvature} \zeta =
\frac{2}{3a^2(1+w)}\left(\frac{\varphi^{(1)}}{a'/a^3}\right )'\,.
\ee
This quantity is extremely useful since for an expanding universe it
is conserved on super-horizon scales. Thus it is the amplitude and
spectrum of these perturbations which are important for the
comparison of theoretical predictions with observations.

We recall that the evolution equation for the first order vector
perturbations is
\[
\nabla^2\dot B^{(v1)}_{\alpha}+2H\nabla^2 B^{(v1)}_\alpha=0\,,
\]
which admits solutions proportional to $1/a^2$.

Now, the right hand side of equation (\ref{vector-identif}) can be
expressed as a gradient of a scalar, hence $v^{(1)}_\alpha$ does not
have a pure vector part, implying that a scalar field does not
support a pure vector velocity perturbation at first order. We
therefore have \be \label{vv1} v^{(v1)}_\alpha=0\,, \ee in this case.
Furthermore, the first order metric vector perturbations satisfy the
momentum constraint equation \be \label{F0V1} \frac{\nabla^2
B^{(v1)}_\alpha}{2a^2}=- 8 \pi G (\rho+p) v^{(v1)}_\alpha \,. \ee In
line with common practice we shall employ Fourier decomposition of
the perturbations. Considering Eq. (\ref{F0V1}) together with Eq.
(\ref{vv1}) implies that $B^{(v1)}_\alpha(k)=0$ for $k\neq0$, where
$B^{(v1)}_\alpha(k)$ are the Fourier modes of the metric vector
perturbations $B^{(v1)}_\alpha$. To proceed, therefore, we fix
$B^{(v1)}_\alpha=0$, so that at first order there are no metric
vector perturbations present.

Finally, to complete our first order analysis we comment on tensor
perturbations. Since we are primarily interested in the evolution of
second order vector perturbations, our interest in first order
tensor perturbations is limited to how their couplings may produce
vector perturbations at second order. However, their effects are
likely to be subdominant as compared to the scalar terms, so we
shall ignore the tensor perturbations at first order. Thus equations
(\ref{Fried})-(\ref{scalar}) together with equation (\ref{FOS1}) and
the expression (\ref{FOS2}) define a closed set of evolution
equations for the background and the first order scalar
perturbations in the Poisson gauge.
%%%%%%%%%%%%%%%%%%%%%%%%%%%%%%%%%%%%%%%%%%%%%%%%%%%%%%%%%%%%%%%%%%%%%%%%%%%%
\subsection{Second order vector perturbations}
\label{secondord}
%%%%%%%%%%%%%%%%%%%%%%%%%%%%%%%%%%%%%%%%%%%%%%%%%%%%%%%%%%%%%%%%%%%%%%%%%
As was mentioned above, an important feature of perturbations at
second and higher orders is that the evolution equations for
scalar, vector and tensor modes
couple. As a result, even in the absence of vector perturbations at
first order, these modes can
be generated at the second order by the scalar-scalar mode couplings as we shall
see below.

Following \cite{NH} (see also  \cite{Nakamura}), and using the
Poisson gauge, the evolution equation for second order vector
perturbations is given by
\begin{eqnarray}
\label{vec-evol}
\frac{1}{2a}\nabla^2(\dot
B^{(v2)}_\alpha+2HB^{(v2)}_\alpha)&=&8 \pi G
\nabla_\beta\Pi^{\beta}_\alpha+ \nabla_\beta N_{4\alpha}^\beta \nonumber \\
&-&\nabla_\alpha\nabla^{-2}\nabla^\gamma\nabla_\beta
(N_{4\gamma}^\beta+8 \pi G\Pi^\beta_\gamma)\,,
\end{eqnarray}
where \be \Pi_{\alpha\beta}=\frac{1}{a^2} \left(
\phi^{(1)}_{,\alpha}\phi^{(1)}_{,\beta}- \frac{1}{3}
\delta_{\alpha\beta}\phi^{(1)}_{,\gamma}\phi^{(1),\gamma}\right)\,,
\ee and \be N^\beta_{4\alpha}=\frac{-1}{2a^2}\left(2\varphi^{(1)}
\varphi^{(1),\beta}_{,\alpha} -\frac{2}{3}
\delta^{\beta}_{\alpha}\varphi^{(1)} \nabla^2\varphi^{(1)} +
\varphi^{(1),\beta}\varphi^{(1)}_{, \alpha} -\frac{1}{3}
\delta^{\beta}_\alpha \varphi^{(1),\gamma} \varphi^{(1)}_{, \gamma}
\right ).\ee Taking $\nabla^2$ of equation (\ref{vec-evol}) we obtain
\begin{eqnarray}
\label{mod-vec-evol}
&~& \frac{1}{4a}\nabla^2\nabla^2(\dot B^{(v2)}_\alpha+2HB^{(v2)}_\alpha) \nonumber \\
&=& 4\pi G\left [\nabla^2\phi^{(1)}\nabla^2\phi^{(1)}_{,\alpha}-
\phi^{(1)}_{,\alpha}\nabla^2\nabla^2\phi^{(1)}+
\phi^{(1)}_{,\beta}\nabla^2\phi^{(1),\beta}_{,\alpha}-\phi^{(1)}_{,\alpha\beta}\nabla^2\phi^{(1),\beta}
\right ] \nonumber\\
&&+2\nabla^2\varphi^{(1)}_{,\beta}\varphi^{(1),\beta}_{,\alpha}-
\varphi^{(1)}_{,\beta}\nabla^2\varphi^{(1),\beta}_{,\alpha}+\varphi^{(1)}_{,\alpha}\nabla^2\nabla^2\varphi^{(1)}\,.
\end{eqnarray}
This equation can be solved by giving $\varphi^{(1)}$, which can be
extracted from the first order evolution equation (\ref{FOS1}),
together with the zeroth order equations. A non-trivial solution to
(\ref{vec-evol}) exists and has the structure \be
\label{solution-vec} B^{(v2)}_\alpha= \frac{C_\alpha({\bf
x})}{a^2}+(\mbox{inhomogeneous terms})\,, \ee where $C_\alpha({\bf
x})$ are arbitrary spatial functions and by ``inhomogeneous
terms" we mean the terms generated by the inhomogeneous part of Eq.
(\ref{mod-vec-evol}). This solution has to be compatible with the
momentum constraint equation, which in the Poisson gauge takes the
particularly simple form:
\begin{equation}
\label{SOV2}
\frac{1}{2a} \nabla^2 B^{(v2)}_{\alpha} = - 8 \pi G a(\rho+p)(v_{\alpha}^{(2)}
-\partial_{\alpha}\nabla^{-2}\partial^\beta v_{\beta}^{(2)})\,,
\end{equation}
where the  second order velocity perturbations, $v^{(2)}_{\alpha}$,
are given by Eq. (\ref{vector-identif2}). We note that, in general,
$v_{\alpha}^{(2)}\ne 0$ and more importantly that
$v_{\alpha}^{(v2)}\ne 0$. This is a crucial difference with respect
to the first order case for which $v_{\alpha}^{(v1)}= 0$ forbids the
existence of first order vector perturbations.

Taking $\nabla^2$ of the equation (\ref{SOV2}) it can be rewritten
using (\ref{vector-identif2}) as
\begin{eqnarray}
\label{vector-eq} \frac{1}{4a}\nabla^2\nabla^2 B^{(v2)}_{\alpha} &=&
(\varphi^{(1),\beta}_{,\alpha} \dot{\varphi}^{(1)}_{,\beta}+
\varphi^{(1)}_{,\alpha}\nabla^2 \dot{\varphi} ^{(1)}-\nabla^2
\varphi^{(1)} \dot{\varphi}^{(1)}_{,\alpha} -
\varphi^{(1)}_{,\beta}\dot{\varphi}^{(1),\beta}_{,\alpha} ) \\&-& 4
\pi G (\nabla^2 \dot{\phi}^{(1)}\phi^{(1)}_{,\alpha}+
\dot{\phi}^{(1)}_{,\beta}\phi^{(1),\beta}_{,\alpha} -
\dot{\phi}^{(1)}_{,\alpha}\nabla^2\phi^{(1)}-
\dot{\phi}^{(1),\beta}_{,\alpha}\phi^{(1)}_{,\beta} )\,. \nonumber
\end{eqnarray}
The right hand side of this expression is in terms of first order
scalar couplings which are in general non-zero. Furthermore, by
direct substitution of Eq. (\ref{vector-eq}) into Eq.
(\ref{mod-vec-evol}), we find that for
(\ref{mod-vec-evol})-(\ref{solution-vec}) and (\ref{vector-eq}) to
be made compatible we must have $C_\alpha({\bf x})=0$.

This demonstrates that, in general, scalar-scalar mode couplings can
produce vector perturbations at second order, even though they are
absent at the first order.  It also demonstrates that these
perturbations are completely determined by the behaviour of the
first order scalar perturbations, since the $C({\bf x})/a^2$
part of the solution (\ref{solution-vec}) must be absent for a scalar
field.

It is also important to recall that there are a number of results
suggesting that vorticity is zero in scalar field settings (see e.g.
\cite{NH,Nakamura}). In our case, this is easy to demonstrate in the
more familiar energy frame using the four velocity associated with
the scalar field
\[
\label{vel} u_a = \frac{1}{N}\phi_{,a}\,,
\]
where $N=|\phi_{,b}\phi^{,b}|^{1/2}$, which implies
\[
\omega_{ab}:=h_{~[a}^c h_{~b]}^d u_{d,c}=0\,,
\]
where $h_{ab}=g_{ab}+u_au_b$.  This is an exact statement and
therefore valid at any perturbation order. So, we have a setting
with non-zero second order vector perturbations and zero vorticity.

Now, to determine the behaviour of second order vector
perturbations concretely, we need to consider particular examples of
contracting universes. We shall do this in the next section.
%%%%%%%%%%%%%%%%%%%%%%%%%%%%%%%%%%%%%%%%%%%%%%%%%%%%%%%%%%%%%%%%%%%%%%%%%%%%%%%%%%%%%%%%%
\section{Analytic solutions}
\label{analytic}
%%%%%%%%%%%%%%%%%%%%%%%%%%%%%%%%%%%%%%%%%%%%%%%%%%%%%%%%%%%%%%%%%%%%%%%%%%%%%%%%%%%%%%%%%%%
\label{power}
In order to calculate the vector perturbations to second order we
shall require solutions to the background and the first order equations.
In general these can be obtained numerically. To proceed
analytically, however, one needs to make assumptions. Concerning the
background equations (\ref{Fried})-(\ref{scalar}), it is well know
{that there exists an  exact solution} if the field is self-interacting through
an exponential potential \cite{PowerLaw}. The result is a power-law
solution in which the scale factor grows with cosmic time as $a =
a_0 t^{r}$. For these solutions $H^2 \propto \dot{\phi}^2 \propto
V(\phi)$ and hence the equation of state, $w = \frac{p}{\rho} =
\frac{\dot{\phi}^2-2V(\phi)}{\dot{\phi}^2+2V(\phi)}$, is a constant.

In both the standard inflationary scenario and scenarios
involving a collapsing
phase {such as the ekpyrotic/cyclic scenario}, it is reasonable
to assume that the
equation of state will be approximately constant for a significant
period of their evolution, and hence the power-law solutions are a
very powerful tool for studying the dynamics of these scenarios
analytically. Moreover, with the background evolving in accordance
with a power-law solution, there is a well known analytic solution
to Eq. (\ref{FOS1}), which then allows the first order scalar
perturbations in this case to be determined analytically \cite{spectrum}.

The form of the potential which gives rise to the power-law
behaviour $a=a_0 t^{r}$ is \be \label{ExpPot} V(\phi) =V_0 e^{
\left(-\sqrt{\frac{16 \pi G}{r}} \right ) \phi }\,, \ee assuming that
$\phi$ is increasing.
%We note that in conformal time
%we have
%$a=a_0^{\frac{1}{1-p}} (-\tau)^{\frac{p}{1-p}}$
%and ${\cal H} \propto \phi' \propto (-\tau)^{-1}$.
Changing to conformal time we have $a=a_0^{\frac{1}{1-r}}
(-\tau)^{\frac{r}{1-r}}$, ${\cal H}=\frac{r}{1-r}\frac{1}{\tau}$ and
$\phi' = \frac{1}{\sqrt{8\pi G}}
\frac{\sqrt{2r}}{1-r}\frac{1}{\tau}$. In the following subsections
we shall only be interested in cases for which $\tau$ is negative and
increasing towards zero, with $-\infty$ and zero representing the
asymptotic past and future respectively. Using conformal time and
taking Fourier transforms allows Eq. (\ref{FOS1}) to be written as
\bea
& & \varphi^{(1)}(k,\tau)''+ 2\left ({\cal H}-
\frac{\phi''}{\phi'}\right)\varphi^{(1)}(k,\tau)' \nonumber \\
&+& k^2
\varphi^{(1)}(k,\tau) + 2\left ({\cal H}'- \cal{H}
\frac{\phi''}{\phi'}\right)\varphi^{(1)}(k,\tau) = 0\,,
\eea
where a
prime denotes differentiation with respect to conformal time $\tau$
and $k=|{\bf k}|$.  This equation can be solved in terms of
Bessel functions. A common procedure is to use the transformation
$\varphi^{(1)}(k,\tau) = \left ( \frac{\phi'}{a} \right ) u(k,\tau)$
to rewrite the above equation in the form
\be
\label{uEq}
u(k,\tau)'' + k^2 u(k,\tau) - \left(\frac{\theta''}{\theta}\right)
u(k,\tau) = 0\,, \ee where $\theta={\cal H}/a\phi'$, and hence
$\frac{\theta''}{\theta} = r/(1-r)^2\tau^{-2}$. The solution to Eq.
(\ref{uEq}) is then readily given by \be
u(k,\tau)=(-\tau)^{\frac{1}{2}} \left( a_k J_\nu( -k\tau)+b_k Y_\nu(
-k\tau) \right) \,,
\ee
where $J_\nu$ and $Y_\nu$ are Bessel
functions of the first and second kind respectively, $\nu=
\frac{1}{2}\left | \frac{1+r}{1-r} \right |$ and $a_k$ and $b_k$ are
arbitrary constants.

In all the scenarios we shall discuss, the perturbations represented
by $u(k,\tau)$ have their origin in
quantum fluctuations which are normalised far inside the
cosmological horizon, but are pushed outside the horizon during the
process of inflation or collapse \cite{Turok}. Inside the horizon this
corresponds to $-k\tau \rightarrow \infty$, and therefore in this
limit we must have $u(k,\tau) \approx
\frac{i}{(2k)^{3/2}}e^{-ik\tau}$ in order to match to the Minkowski
vacuum. This limit allows the free constants to be fixed and we
arrive at \be \label{u} u(k,\tau)=
\frac{\sqrt{\pi}i}{4k}e^{i(\nu+\frac{1}{2})\frac{\pi}{2}}\sqrt{-\tau}
H_\nu(-k\tau)\,, \ee where $H_\nu=J_\nu(x) + i Y_\nu(x)$ is a Hankel
function. The limit corresponding to the modes pushed outside the
horizon is given by $-k\tau \rightarrow 0$. Expanding the Hankel
function in this limit using
\begin{eqnarray}
J_{\nu}(x) &=& \frac{1}{\Gamma(\nu +1)}
\left(\frac{x}{2}\right)^{\nu}-\frac{1}{\Gamma(\nu+2)}\left(\frac{x}{2}\right)^{\nu+2}+O(x^{\nu+4})\,, \\
Y_{\nu}(x) &=& -\frac{1}{\pi}
\left(\frac{x}{2}\right)^{-\nu}\left(\Gamma(\nu)+\frac{\Gamma(\nu-1)}{4}x^2\right)+O(x^{4-\nu})\,,
\nonumber
\end{eqnarray}
we arrive at
\begin{eqnarray}
\label{cphi} \varphi^{(1)}(k,\tau) & = & C_r
k^{\nu-1}(-\tau)^{\nu-\frac{1}{2}-\frac{r}{1-r}}+ D_r
k^{\nu+1}(-\tau)^{\nu+\frac{3}{2}-\frac{r}{1-r}}
+O(\tau^{\nu+3-\frac{r}{1-r}}) \nonumber \\
&&+B_r
k^{-\nu+1}(-\tau)^{-\nu+\frac{3}{2}-\frac{r}{1-r}}
+A_r k^{-\nu-1}(-\tau)^{-\nu-\frac{1}{2}-\frac{r}{1-r}} \nonumber \\
& & +O(\tau^{3-\nu-\frac{r}{1-r}})\,,
\end{eqnarray}
where $A_r={\cal K}_r\Gamma(\nu),~~B_r={\cal
K}_r\frac{1}{4}\Gamma(\nu-1),~~C_r={\cal K}_r \frac{i\pi}
{\Gamma(\nu-1)}, D_r=-{\cal K}_r\frac{i\pi }{4\Gamma(\nu+2)}$, with
${\cal K}_r=
\frac{\sqrt{r}}{1-r}\frac{1}{a_0^{\frac{1}{1-r}}}\left(2^{\nu-\frac{7}{2}}
e^{i(\nu+\frac{1}{2})\frac{\pi}{2}}\frac{1}{\sqrt{\pi}\sqrt{8\pi
G}}\right)$, are constants which are related to each other and
depend on the initial data. This feature will play an important role in what
follows.

With the exact solution for $\varphi^{(1)}$ at hand, we are now able
to evaluate the second order vector perturbations $B^{(v2)}_\alpha$
from Eq. (\ref{vector-eq}), for a universe undergoing power--law
evolution. A potentially important indicator to differentiate
between collapsing models and expanding inflationary models could be
sought in the ratio of the amplitudes of $B^{(v2)}_\alpha$ between
the expanding and contracting phases.
%%%%%%%%%%%%%%%%%%%%%%%%%%%%%%%%%%%%%%%%%%%%%%%%%%%%%%%%%%%%%%%%%%%%%%%%%%%%%%%%%%%%%%%%%%%%%%
\section{Amplitudes of vector perturbations in
expanding and contracting phases}
%%%%%%%%%%%%%%%%%%%%%%%%%%%%%%%%%%%%%%%%%%%%%%%%%%%%%%%%%%%%%%%%%%%%%%%%%%%%%%%%%%%%%%%%%%
\label{Bcalc}

In this subsection we evaluate the amplitudes of $B^{(v2)}_\alpha$
for the expanding and contracting phases. We restrict our attention
to three scenarios which can be well modelled by the power-law
solution introduced above. All these models can produce a nearly
scale--invariant spectrum of first order scalar perturbations in a
straightforward manner. The first order curvature perturbations $\zeta$
must have a spectrum of this form in order to be compatible with
observations. Thus the scale--invariance cannot be used to
distinguish between these scenarios. We shall instead study whether
the resulting amplitudes of the second order vector perturbations
can be used to distinguish between these models.

The first model we shall consider, $r \to \infty$, corresponds to
the standard expanding inflationary scenario. As we shall see, this
limit of $r$ gives rise to scale-invariant spectra for both
$\varphi^{(1)}$ and curvature perturbations. The second model we
shall consider is the ekpyrotic/cyclic model, in which a contracting
phase with $r \to 0$ replaces the inflationary epoch. This limit of
$r$ also gives rise to a scale-invariant spectrum for the
$\varphi^{(1)}$ perturbations, but not for the curvature
perturbations $\zeta$. In the ekpyrotic/cyclic scenario part of the
$\varphi^{(1)}$ perturbation is matched onto the curvature
perturbation when the universe undergoes a bounce, and hence the
observationally important quantity $\zeta$ has a scale-invariant
spectrum after the bounce \cite{Tolley:2003nx,McFadden:2005mq}.
Finally, we shall consider a dust-like contracting universe
\cite{Wands,Finelli,AllenWands}, which corresponds to a contracting
universe with $r \to 2/3$.  While it has been pointed out that,
unlike the previous two cases, this solution is not a dynamical
attractor \cite{HeardWands,Turok}, it is still of considerable
interest since $r = 2/3$ gives rise to a scale-invariant spectrum
for the curvature perturbations.  We note that the limiting values
of $r$ we have discussed here are of course idealised. In a
realistic setting $r$ would approach but not reach such limits. This
is important to keep in mind as the solutions for $r=0$ and
$r=\infty$ are clearly ill defined.

A comparison of these three cases is given in Ref. \cite{Turok}. We
shall now consider the three cases in turn, and evaluate
$B^{(v2)}_\alpha$ for each. We shall then proceed to calculate the
ratios between the amplitudes of the second order perturbations
$B^{(v2)}_\alpha$ in the inflationary case and each of the
contracting cases respectively.

In what follows it is
useful to define the constant quantities
\[
\gamma_r=\frac{r}{1-r}~~~\mbox{and}~~~\chi_r=\frac{1}{\sqrt{8\pi
G}}\frac{\sqrt{2r}}{1-r}\,.
\]
%%%%%%%%%%%%%%%%%%%%%%%%%%%%%%%%%%%%%%%%%%%%%%%%%%%%%%%%%%%%%%%%%%%%%%%%%%%%%%%%%%
\subsection{Expanding inflationary phase ($r\to\infty$):}
%%%%%%%%%%%%%%%%%%%%%%%%%%%%%%%%%%%%%%%%%%%%%%%%%%%%%%%%%%%%%%%%%%%%%%%%%%%%%%%
This expanding phase is characterised by $r\to\infty$ (and
$\nu=\frac{1}{2}$) \cite{Turok}. Substituting in (\ref{cphi}) we
obtain \be \label{scalar-inflation} \varphi(k,\tau)= A_\infty
k^{-\frac{3}{2}} + C_\infty
 k^{-\frac{1}{2}}(-\tau)+ B_\infty k^{\frac{1}{2}} (-\tau)^2+O(\tau^3)\,.
\ee For a perturbation to have a scale-invariant spectrum its
Fourier components must be proportional to $k^{-3/2}$. Thus
considering the expression above, we recover the well known result
that as $\tau\to 0$, inflation produces a scale-invariant power
spectrum for linear scalar perturbations.

To evaluate the amplitude of the second order vector perturbations
we consider the expression (\ref{vector-eq}). To proceed we shall
employ the property that the Fourier transform $\hat\varphi(k,
\tau)= \int \varphi({\bf x},\tau) e^{-i{\bf k}\cdot{\bf x}} d{\bf
x}$ of the product $\varphi_1({\bf x},\tau)\varphi_2({\bf x},\tau)$
is equal to the convolution
$\hat\varphi_1(k_1,\tau)\star\hat\varphi_2(k_2,\tau)=\int\hat\varphi_1
(k_2,\tau)\hat\varphi_2 (k_1-k_2,\tau) d{\bf k_2}$. So e.g. $ \int
(\nabla^2
\varphi\dot\varphi_{,\alpha}-\varphi_{,\alpha}\nabla^2\dot\varphi)
e^{-i{\bf k}\cdot{\bf x}} d{\bf x}=i\int {\bf k}_2^{2}({\bf
k}_1-{\bf k}_2)(\hat\varphi_{k_2}\dot{\hat\varphi}_{
k_1-k_2}-\dot{\hat\varphi}_{k_2}\hat\varphi_{k_1-k_2})d{\bf k_2}$.
Then, by substituting (\ref{scalar-inflation}) into
(\ref{vector-eq}) and using the Fourier transform we find
\begin{eqnarray}
k_1^4 {\bf B}^{(v2)} &=& 4i\left(1-\frac{\gamma_\infty(\gamma_\infty
+1)}{4\pi G\chi_\infty^2}\right)A_\infty C_\infty \nonumber \\
&\times& \int d^3k_2{\bf
F}
({\bf k_1,k_2})(a_{k_2}c_{k_1-k_2}-c_{k_2}a_{k_1-k_2})\nonumber\\
&+&O(\tau)\,,
\end{eqnarray}
where ${\bf F}({\bf k_1,k_2})={\bf k}_2^2 ({\bf k}_1-{\bf k}_2)-[{\bf k}_2\cdot({\bf k}_1-{\bf k}_2)]{\bf k}_2$,
$a_{k}={k}^{-\frac{3}{2}}$, $b_{k}={k}^{\frac{1}{2}}$ and
$c_k={k}^{-\frac{1}{2}}$, with $k=|{\bf k}|$.
%%%%%%%%%%%%%%%%%%%%%%%%%%%%%%%%%%%%%%%%%%%%%%%%%%%%%%%%%%%%%%%%%%%%%%%%%%%%%%%%%%
\subsection{Recollapsing phase ($r \to 0$):}
%%%%%%%%%%%%%%%%%%%%%%%%%%%%%%%%%%%%%%%%%%%%%%%%%%%%%%%%%%%%%%%%%%%%%%%%%%%%%%%
\label{r0} This recollapsing phase is characterised by $r\to 0$ (and
$\nu=\frac{1}{2}$) \cite{Turok}. Substituting in (\ref{cphi}) we
obtain \be \label{scalar-recollapse1} \varphi^{(1)}(k,\tau)= A_0
k^{-\frac{3}{2}} (-\tau)^{-1}+C_0 k^{-\frac{1}{2}}+ B_0
k^{\frac{1}{2}} (-\tau) +O(\tau^2)\,. \ee Thus we recover the well
known result that the linear scalar perturbations grow (diverge) in
a collapsing phase if the so-called decaying modes are taken into
account. This in turn implies (from (\ref{vector-eq})) that the
amplitudes of second order vector perturbations also grow during a
collapsing phase, as the inverse square of the conformal time. We
can also see that $\varphi^{(1)}$ has a scale-invariant spectrum in
the $r \to 0$ limit.

In order to analyse this behaviour in more detail we substitute
(\ref{scalar-recollapse1}) into (\ref{vector-eq}) and use the
Fourier transform to obtain
\begin{eqnarray}
{k_1}^4 {\bf B}^{(v2)} &=& 4i\left(1-\frac{\gamma_0(\gamma_0
-1)}{4\pi G\chi_0^2}\right)\frac{A_0C_0}{\tau^2} \nonumber \\
&\times& \int d^3k_2{\bf F}({\bf k_1,k_2})(c_{k_2}a_{k_1-k_2}-a_{k_2}c_{k_1-k_2})\nonumber\\
&+&O(1)\,.
\end{eqnarray}
It is interesting to note that the form of Eq. (\ref{vector-eq})
leads to a cancellation which results in the leading order time
dependence in this case to be proportional to $\tau^{-2}$ rather
than $\tau^{-3}$ as might be expected.
%%%%%%%%%%%%%%%%%%%%%%%%%%%%%%%%%%%%%%%%%%%%%%%%%%%%%%%%%%%%%%%%%%%%%%%%%%%%%%%%%%
\subsection{Recollapsing phase ($r=\frac{2}{3}$):}
%%%%%%%%%%%%%%%%%%%%%%%%%%%%%%%%%%%%%%%%%%%%%%%%%%%%%%%%%%%%%%%%%%%%%%%%%%%%%%%
This dust-like collapsing phase is characterised by $r=2/3$ (and
$\nu=5/2$) \cite{Wands,Finelli}.  Substituting in (\ref{cphi}) we
obtain \be \varphi^{(1)} (k,\tau)= A_{\frac{2}{3}}
k^{-\frac{7}{2}}(-\tau)^{-5} + B_{\frac{2}{3}} k^{-\frac{3}{2}}
(-\tau)^{-3}+ O(\tau^{-\frac{3}{2}}) \ee Again we find that the
scalar perturbations $\varphi^{(1)} (k,\tau)$ grow (diverge) in the
collapsing approach to the singularity, but at a substantially
different rate from the previous collapsing scenario. We shall see
this has important consequences for the analysis of the next
section. The second order vector perturbations in this case become
\begin{eqnarray}
{k_1}^4 {\bf B}^{(v2)} &=&
8i\left(4-\frac{4\gamma^2_{\frac{2}{3}}-17\gamma_{\frac{2}{3}}+75}{4\pi
G\chi_{\frac{2}{3}}^2}\right)\frac{A_{\frac{2}{3}}B_{\frac{2}{3}}}{(-\tau)^9}
\nonumber \\
&\times& \int
d^3k_2{\bf F}({\bf k_1,k_2})
(b_{k_2}a_{k_1-k_2}-a_{k_2}b_{k_1-k_2})\nonumber\\
&+&O(\tau^{-4})\,.
\end{eqnarray}
%%%%%%%%%%%%%%%%%%%%%%%%%%%%%%%%%%%%%%%%%%%%%%%%%%%%%%%%%%%%%%%%%%%%%%%%%%%%%%%%%%
\subsection{Comparison of expanding and recollapsing phases}
%%%%%%%%%%%%%%%%%%%%%%%%%%%%%%%%%%%%%%%%%%%%%%%%%%%%%%%%%%%%%%%%%%%%%%%%%%%%%%%
As discussed above, a potentially important signature of a
collapsing phase can be obtained by comparing the ratio between the
amplitudes of the second order metric vector perturbations of the
collapsing and expanding phases. In this subsection we shall
evaluate this ratio for both contracting scenarios.

Comparing the ekpyrotic collapsing ($r\to 0$) and the inflationary
($r\to \infty$) cases we find the ratio of the amplitudes of the
second order vector perturbations in these cases to be
\be
\label{ratio1} \frac{ |{\bf B}^{(v2)coll}|}{|{\bf B}^{(v2)exp}|}=
K_1 \frac{A_0C_0}{A_\infty C_\infty}\frac{1}{\tau_{coll}^2}+O(1)\,,
\ee
where $K_1$ is a constant in time, which shows that
this ratio grows with the duration of the collapse phase as
$1/\tau_{coll}^2$.

Comparing the dust-like collapsing ($r=2/3$) and the inflationary
($r\to \infty$) cases we find the ratio of the amplitudes of the
second order vector perturbations in these cases to be
\be
\label{ratio3} \frac{| {\bf B}^{(v2)coll}|}{ |{\bf B}^{(v2)exp}|}
=K_3\frac{A_{\frac{2}{3}}B_{\frac{2}{3}}}{A_\infty
C_\infty}\frac{1}{\tau_{coll}^9} +O(\tau^{-4}_{coll})\,,
\ee
where
$K_3$ is a constant in time, which shows that the ratio in this case
is radically different from the ekpyrotic case, growing with the
duration of the collapse phase as $1/\tau_{coll}^9$.

Despite their usefulness these expressions are not, as they stand,
sufficient to provide the complete information required for
observational purposes. This is because we have not taken into account
the constraints on the amplitude of the first order scalar
perturbations. We shall proceed to implement these constraints in
the next section.
%%%%%%%%%%%%%%%%%%%%%%%%%%%%%%%%%%%%%%%%%%%%%%%%
\subsubsection{Effects of imposing observational constraints from first order
curvature perturbations:}
%%%%%%%%%%%%%%%%%%%%%%%%%%%%%%%%%%%%%%%%%%%%%%
We have seen that in the scenarios which involve a collapsing phase
the amplitude of the first order scalar perturbations
$\varphi^{(1)}$ grows as the collapse proceeds.  At the end of the
collapse, however, the first order perturbations must have the
correct amplitude to be compatible with observations. It is
therefore important to study the consequences of demanding that at
the end of the collapse the first order scalar perturbations
produced are equal to those obtained from observations.

As was mentioned above, for expanding universes, and hence for
inflation, the curvature perturbation on comoving hypersurfaces is
conserved on super-horizon scales. This quantity is therefore used
for comparison between theory and observations. In cosmologies
with a collapsing phase we require the value of the curvature
perturbation at the wavenumber corresponding to the largest
scale on the CMB after the bounce to be equal to the required
observational value, which is in turn equal to the value produced by
a successful inflationary model. We note that since this study
is only concerned with ratios we do not need to give the
required observational value explicitly, and since the three
scenarios considered here all produce the same spectral
dependence we do not need to specify the wavenumber explicitly.

For the scaling scenarios considered here we have
\[
\frac{a'}{a}=\left(\frac{r}{1-r}\right)\frac{1}{\tau}\,~,~~~~~~~~~
~\frac{a''}{a'}=\left(\frac{r}{1-r}-1\right)\frac{1}{\tau}\,.
\]
Thus in the expanding phase $r\to \infty$ we find, using
Eq. (\ref{Curvature}), that
\[
\zeta_{exp}=\frac{2}{3}\frac{1}{1+w_{exp}}\left(C_\infty
k^{-\frac{3}{2}} -B_\infty k^{\frac{1}{2}}(-\tau)^2+O(\tau^3)\right)\,,
\]
while in the collapsing case $r\to 0$ we obtain
\be
\label{coll1-ex}
\zeta_{coll1}=\frac{2}{3}\frac{1}{1+w_{coll1}}\left(\frac{1-r}{r}\right)\left(C_0
k^{-\frac{1}{2}}-2B_0 k^{\frac{1}{2}} (-\tau) +O(\tau^2)\right)\,.
\ee
Finally in the $r\to 2/3$ collapse case we find
\be
\label{coll2-ex}
\zeta_{coll2}=\frac{2}{3}\frac{1}{1+w_{coll2}}\left(B_{\frac{2}{3}}k^{-\frac{3}{2}}(-\tau)^{-3}
+O(\tau^{-\frac{3}{2}}) \right)\,.
\ee
Considering the expression (\ref{coll2-ex}) for $\zeta_{coll2}$,
we recover the result that the collapsing  scenario with a dust
like equation of state produces scale-invariant curvature perturbations.

In the ekpyrotic scenario $\varphi^{(1)}$ is responsible for the
curvature perturbation after the bounce. The argument used is that
$\varphi^{(1)}$ and $\zeta$ mix at the bounce and part of
$\varphi^{(1)}$ at the end of the collapse is matched onto the
curvature perturbations after the bounce. We therefore equate, at the
lowest order, the perturbations
\[
\zeta_{exp}=\psi \varphi_{coll1}\,,
\]
where $\psi$ represents the proportionality factor between the two
phases in the ekpyrotic scenario \cite{Tolley:2003nx, Steinhardt:2004gk}. This
results in the constraint
\[
C_\infty=\frac{3}{2}(1+w_{exp})A_0\psi\frac{1}{\tau_{coll1}}\,,
\]
which can be used to obtain the ratio between the amplitudes of the
second order vector perturbations in the collapsing and expanding
phases from (\ref{ratio1})
\begin{equation}
\frac{|{\bf B}^{(v2)coll1}|}{| {\bf
B}^{(v2)exp}|}=K_1\left(\frac{4/3}{(1+w_{exp})\psi}\right)^2+O(\tau_{coll1}^2)\,.
\end{equation}
This is a constant at the lowest order (independent of the length of
the collapsing phase) thus indicating that the important factor
determining the ratio is the proportionality factor $\psi$ in the
ekpyrotic scenario, which is determined by the physics of the
bounce.

In the case of $r\to 2/3$ collapsing scenario, however, the
curvature perturbations survive through the bounce and are
thereafter conserved. Hence at the lowest order we can equate
\[
\zeta_{exp}=\zeta_{coll2}\,,
\]
which implies
\[
C_\infty= B_{\frac{2}{3}}\left(\frac{1+w_{exp}}{1+w_{coll2}}\right)
\frac{1}{(-\tau)^{3}}\,,
\]
and from (\ref{ratio3})
\begin{equation}
\frac{|{\bf B}^{(v2)coll2}|}{|{\bf B}^{(v2)exp}|}
=12K_3\left(\frac{1+w_{coll2}}{1+w_{exp}}\right)^2
\frac{1}{\tau_{coll2}^3}+O(\tau^{2}_{coll2})\,.
\end{equation}
Therefore, at the lowest order, the ratio of the amplitudes of the
second order vector perturbations is proportional to
$\tau_{coll}^{-3}$, which increases as the collapse time tends to
$0$.  Hence for a long collapsing phase we expect this ratio to
become very large.
%%%%%%%%%%%%%%%%%%%%%%%%%%%%%%%%%%%%%%%%%%%%%%%%%%%%%%%%%%%%%%%%%%%%%%%%%%%
\section{Conclusions}
\label{conclusion}
%%%%%%%%%%%%%%%%%%%%%%%%%%%%%%%%%%%%%%%%%%%%%%%%%%%%%%%%%%%%%%%%%%%%%%%%%%%
A great deal of effort has recently gone into constructing alternative
models to the standard inflationary scenario motivated by
developments in string/M-theory. Many of these models involve a
contracting phase. As a step towards distinguishing these models
from the standard inflationary scenario we have studied the
generation and evolution of vector perturbations in collapsing
phases, in the context of scalar field cosmologies.

There are a number of reasons why vector perturbations might provide
useful signatures in models with contracting phases. Such
perturbations are highly suppressed by inflation and are extremely
difficult to generate at the early epochs after inflation. However,
they would grow if produced during collapsing phases.

Noting that first order vector perturbations cannot be supported by
regimes purely sourced by a scalar field, we have considered second
order vector perturbations. We have found that such perturbations
can be generated by mode-mode couplings of the first order scalar
perturbations and have derived their explicit dependence on first
order perturbations. In principle our expressions allow
the spectral dependence of second order vector perturbations
to be determined, though in this work we have focused on their temporal
behaviours in various early universe scenarios.
Considering exponential potentials, which allow
power-law solutions, we have studied the ratio of the amplitudes of
second order vector perturbations in contracting and expanding
phases. We have found that ignoring the details of how the hot big
bang is recovered, the relative magnitudes of the second order
vector perturbations in the two phases depend on the scaling
solutions chosen. In particular we have considered two collapsing
models, the first motivated by the ekpyrotic/cyclic scenario and the
second by the dust-like collapsing scenario given in Ref.
\cite{Wands,Finelli}.

For the first case, we have found the ratio to be independent of the
length of collapse, at the lowest order, while depending on the
details of the matching of the perturbations at the bounce. This is
counter to the expectation that vector perturbations should be more
prominent after a longer collapsing phase.  This result is a
consequence of the fact that the vector perturbations in this case
 grow as the square of the rate at which the scalar
perturbations $\varphi^{(1)}$ grow,  which in turn is a consequence
of the cancellation mentioned in section \ref{r0}. In this case,
$\varphi^{(1)}$ is responsible for the curvature perturbations after
the bounce. Hence, once the observational constraint, namely that
the curvature perturbations produced by this collapsing phase must
have the same amplitude as the curvature  perturbations produced by
inflation, is imposed, the ratio of vector perturbations becomes
fixed, independently of the collapse time.

For the second case, we find that the ratio increases with the
length of the collapse, being proportional to ${\tau_{coll}}^{-3}$.
This result is due to the fact that the second order vector
perturbations grow more rapidly during the collapse than the square
of the first order curvature perturbations. This implies that the
magnitude of vector perturbation is no longer limited to being of
the order of the square of the first order perturbations, but could
grow to be larger. Hence, even when the constraint on the first
order curvature perturbations is imposed, there is still a
dependence on the collapse time.  This implies that the amount of
vector perturbations present at the end of the collapsing phase
could, in principal, be much larger than that present at the end of
the inflationary epoch, since $\tau \to 0$ as the collapse proceeds.

It is important to note that the amplitude of the vector
perturbations, $B^{(v2)}_\alpha$, calculated here is only valid if the
power-law solutions of section \ref{power} are applicable. These
solutions are clearly only approximate and will not be valid through
the entire inflationary or collapsing phases considered here. In
particular, these behaviours must break down as the universe exits
these phases and enters the hot big bang phase of evolution.  In the
inflationary scenario it is usual to assume that this occurs when
the potential evolves such that it no longer gives rise to
accelerated expansion, after which reheating occurs.  In the
collapsing scenarios the power--law evolution is expected to break
down in the vicinity of the bounce, which is believed to lead
to a radiation dominated phase. In principal, therefore, we would
like to know how vector perturbations are affected by these
transitions. This would, however, almost certainly require a
numerical investigation, beyond the scope of the present study.

In conclusion, a complete understanding of second order vector
perturbations from a collapsing phase and their possible
observational consequences would require a more detailed
understanding of how such perturbations propagate between the scalar
field phase and the hot big bang phase. However, by using power-law
solutions we can compare vector perturbations produced in collapsing
scenarios with those obtained in the inflationary scenario at a
point immediately before the transition into the hot big bang phase.
Our study of the two collapsing scenarios indicates that the
observable differences between the collapsing models and the
inflationary scenario could be large, particularly for the
dust--like collapse, if we assume that the transition we have just
discussed does not significantly alter the ratios we have
calculated.

\section*{Acknowledgements}
We would like to thank Roy Maartens, Karim Malik, Kouji
Nakamura and David Wands for helpful discussions. FCM is supported
by CMAT - Univ. Minho and FCT (Portugal) through grant
SFRH/BPD/12137/2003. DJM is supported by the Centre for Theoretical
Cosmology, Univ. Cambridge. FCM and RT thank TIFR and IUCAA, India,
for kind hospitality where part of this work was done. FCM and DJM
thank the Astronomy Unit, Queen Mary, Univ. London, for hospitality
while most of this work was done.

%%%%%%%%%%%%%%%%%%%%%%%%%%%%%%%%%%%%%%%%%%%%%%%%%%%%%%%%%%%%%%%%%%%%%%%%%%
%%%%%%%%%%%%%%%%%%%%%%%%%%%%%%%%%%%%%%%%%%%%%%%%%%%%%%%%%%%%%%%%%%%%%%%%%

\section*{References}

%\bibliographystyle{JHEPmod}
%\bibliography{paper.bib}
\providecommand{\href}[2]{#2}\begingroup\raggedright\endgroup
\end{document}